\documentclass[%
preprint,
 amsmath,amssymb,
 aps,
longbibliography]{revtex4-1}

\usepackage[abs]{overpic}
\usepackage{graphicx}% Include figure files
\usepackage{dcolumn}% Align table columns on decimal point
\usepackage{bm}% bold math
\usepackage{placeins}
\usepackage{subfig}
\usepackage{color}
\usepackage[normalem]{ulem}
\begin{document}

%\preprint{APS/123-QED} 

\title{Do fluid particles separate exponentially in the dissipation range?}


\author{Rohit Dhariwal}
\author{Andrew D. Bragg}
\email{andrew.bragg@duke.edu}
\affiliation{Department of Civil and Environmental Engineering, Duke University, Durham, North Carolina, USA}

\date{\today}% It is always \today, today,
       % but any date may be explicitly specified

\begin{abstract}
In this paper we consider how the statistical moments of the separation between two fluid particles grow with time when their separation lies in the dissipation range of turbulence. In this range the fluid velocity field varies smoothly, and the relative velocity of two fluid particles depends linearly upon their separation. While this may suggest that the rate at which fluid particles separate is exponential in time, this is not guaranteed because the strain-rate governing their separation is a strongly fluctuating quantity in turbulence. Indeed, the recent paper by Afik \& Steinberg (Nat. Commun. \textbf{8}: 468, 2017) argues that there is no convincing evidence that the moments of the separation between fluid particles grow exponentially with time in the dissipation range of turbulence. Motivated by this, we use Direct Numerical Simulations (DNS) to compute the moments of the particle separation over very long periods of time to see if we ever see evidence for exponential separation. Our results show that if the initial separation between the particles is infinitesimal, the moments of the particle separation first grow as power laws in time, but we then observe, for the first time, convincing evidence that at sufficiently long times the moments do grow exponentially. However, this exponential growth is only observed after extremely long times $\gtrsim 200\tau_\eta$, where $\tau_\eta$ is the Kolmogorov timescale. This is due to fluctuations in the strain-rate about its mean value measured along the particle trajectories, the effect of which on the moments of the particle separation persists for very long times. We also consider the Backward-in-Time (BIT) moments of the article separation, and observe that they too grow exponentially in the long-time regime. However, a dramatic consequence of the exponential separation is that at long-times the difference between the rate of the particle separation Forward-in-Time (FIT) and BIT grows exponentially in time, leading to incredibly strong irreversibility in the dispersion. This is in striking contrast to the irreversibility of their relative dispersion in the inertial range, where the difference between FIT and BIT is constant in time according to Richardson's phenomenology.
\end{abstract}

\maketitle

\section{Introduction}
The problem of the relative dispersion of fluid (tracer) particles in turbulence is an old problem, yet one that remains a very active area of research due to a number of open questions \cite{salazar09}. Some of the seminal studies on the topic are those by Richardson \cite{richardson26} and Batchelor \cite{batchelor52b}. To state their key results, we must first introduce some notation. Let $\bm{r}(t)$ be the vector describing the time-dependent separation between two particles, and $r(t)\equiv \|\bm{r}(t)\|$. The mean-square separation of the two particles may then be written as $\langle {r}^2(t)\rangle_\xi$, where $\langle\cdot\rangle_\xi$ denotes an ensemble average conditioned on the particle-pair satisfying ${r}(0)=\xi$. Note that $t=0$ does not necessarily correspond to the initial time of the whole flow. Rather, it simply denotes the time at which the fluid particles are ``marked''.

Let us first consider initial separations in the inertial range, $\eta\ll\xi\ll\ell$, where $\eta$ is the Kolmogorov length scale and $\ell$ is the integral length scale. Batchelor predicted that at short-times the separation should grow ballistically 
\begin{align}
\Big\langle {r}^2(t)\Big\rangle_\xi=\xi^2+ \Big\langle\|\Delta\bm{u}(\xi,0)\|^2\Big\rangle  t^2,\quad\text{for}\, t\ll\tau_\xi,\label{Bmss}
\end{align}
where $\Delta\bm{u}(\xi,0)$ is the difference in the velocity of the two fluid particles at the initial time, and $\tau_\xi$ is the eddy turnover timescale for an eddy of size $\xi$. In a sufficiently high Reynolds number turbulent flow, when $t\gg \tau_\xi$ the separation between the two fluid particles will remain in the inertial range of the turbulence, i.e. $\eta\ll r(t) \ll\ell$ . For this regime, Richardson predicted 
\begin{align}
\Big\langle {r}^2(t)\Big\rangle_\xi=\mathfrak{g}^F\langle\epsilon\rangle t^3,\quad\text{for}\, t\gg \tau_\xi,\label{Rlaw}
\end{align}
where $\langle\epsilon\rangle$ is the mean kinetic energy dissipation rate of the turbulent flow, and $\mathfrak{g}^F$ is the non-dimensional Richardson constant, measured in experiments to be $\approx 0.55$ \cite{berg06a}. Equation \eqref{Rlaw} describes a super-diffusive growth of the separation between the two fluid particles in the inertial range, with dramatic consequences for understanding the rates at which fluid particles disperse in turbulent flows (detailed discussions of these results can be found in \cite{sawford01,falkovich01,salazar09}).

Whereas the Batchelor prediction \eqref{Bmss} has been confirmed in numerous experimental and numerical studies \cite{yeung04,ouellette06c,bec10b,buaria15,bragg16}, Richardson's prediction \eqref{Rlaw} is notoriously difficult to observe. One of the main reasons for this is that very large Reynolds numbers are required in order for $\eta\ll r(t) \ll\ell$ to be satisfied when $t\gg \tau_\xi$, since $\ell/\eta \sim Re^{3/4}$ (e.g. Pope \cite{pope}). Several quite recent experimental and numerical studies \cite{bitane12,bitane13,ni13,buaria15,bragg16} lend support for \eqref{Rlaw}, although the results are not conclusive. 

All of the aforementioned results refer to Forward-in-Time (FIT) dispersion. The Backward-in-Time (BIT) dispersion is also very important, especially for quantifying particle and scalar mixing in turbulence \cite{corrsin52,durbin80,thomson03,sawford05}. For BIT the mean-square separation is denoted by $\langle {r}^2(-t)\rangle_\xi$, and describes the separation of particle-pairs at earlier times that satisfy the terminal condition $r(0)=\xi$. Under the same arguments invoked by Batchelor and Richardson, \eqref{Bmss} also describes $\langle {r}^2(-t)\rangle_\xi$ in the regime $t\ll \tau_\xi$, and \eqref{Rlaw} describes $\langle {r}^2(-t)\rangle_\xi$ except that $\mathfrak{g}^F$ is now replaced with the BIT constant $\mathfrak{g}^B$, and $\mathfrak{g}^B>\mathfrak{g}^F$ in 3-dimensional turbulence \cite{buaria15,bragg16}.

For $\xi$ in the dissipation range, the frequently quoted result is that the mean-square separation grows exponentially $\langle{r}^2(t)\rangle_\xi= \xi^2 e^{\mathcal{A}^Ft/\tau_\eta}$ \cite{sawford01,salazar09,ni13,costa17}, where $\mathcal{A}^F$ is a non-dimensional number and $\tau_\eta$ is the Kolmogorov timescale. The same arguments leading to $\langle{r}^2(t)\rangle_\xi= \xi^2 e^{\mathcal{A}^Ft/\tau_\eta}$ would also imply $\langle {r}^2(-t)\rangle_\xi=\xi^2 e^{\mathcal{A}^B t/\tau_\eta}$, where we would expect $\mathcal{A}^B\neq \mathcal{A}^F$ due to irreversibility in the dispersion.

Although the result $\langle{r}^2(t)\rangle_\xi= \xi^2 e^{\mathcal{A}^Ft/\tau_\eta}$ is sometimes presented as if it were valid for arbitrary $t$ (e.g. \cite{salazar09,ni13,costa17}), Batchelor \cite{batchelor52a} only claimed that this result applies for $t\gg\tau_\eta$, as we shall explain in more detail later. Furthermore, despite the widespread belief in the validity of $\langle{r}^2(t)\rangle_\xi= \xi^2 e^{\mathcal{A}^Ft/\tau_\eta}$, the evidence for it is very limited, and as we discuss in \S\ref{Evid}, the evidence that has been presented for it is questionable. Indeed, the recent paper by Afik \& Steinberg \cite{afik17} presents a literature review on the topic, upon the basis of which they conclude that there exists no convincing evidence that $\langle{r}^2(t)\rangle_\xi= \xi^2 e^{\mathcal{A}^Ft/\tau_\eta}$ describes how fluid particle-pairs separate in the dissipation range of turbulent flows. The same conclusions also apply to the BIT result $\langle {r}^2(-t)\rangle_\xi=\xi^2 e^{\mathcal{A}^B t/\tau_\eta}$.

In view of these considerations, the purpose of this paper is first to consider the theoretical basis for $\langle{r}^2(t)\rangle_\xi= \xi^2 e^{\mathcal{A}^Ft/\tau_\eta}$ in \S\ref{PPD}, and to show that it is not valid for arbitrary $t$. Then in \S\ref{Evid} we present a literature review to consider claims that have been made to validate $\langle{r}^2(t)\rangle_\xi= \xi^2 e^{\mathcal{A}^Ft/\tau_\eta}$, discussing various problems with these claims. We then introduce in \S\ref{test} a way to conclusively test $\langle{r}^2(t)\rangle_\xi= \xi^2 e^{\mathcal{A}^Ft/\tau_\eta}$, and in \S\ref{RD} we present results from a Direct Numerical Simulation (DNS) study that provide, for the first time, convincing support for $\langle{r}^2(t)\rangle_\xi= \xi^2 e^{\mathcal{A}^Ft/\tau_\eta}$ in the regime $t\gg\tau_\eta$, and also for the more general case $\langle{r}^N(t)\rangle_\xi$ having a similar exponential form in the long-time regime.
\section{Pair-dispersion in the dissipation range}\label{PPD}
In the dissipation range of turbulence, the fluid velocity field $\bm{u}(\bm{x},t)$ varies smoothly in space, and we have
\begin{align}
\Delta\bm{u}(\bm{x},\bm{r},t)=\bm{\Gamma}(\bm{x},t)\bm{\cdot r}+\mathcal{O}(\|\bm{r}\|^2),\label{llff}
\end{align}
where $\Delta\bm{u}(\bm{x},\bm{r},t)\equiv\bm{u}(\bm{x}+\bm{r},t)-\bm{u}(\bm{x},t)$ is the fluid velocity increment, and $\bm{\Gamma}(\bm{x},t)\equiv\bm{\nabla u}(\bm{x},t)$. If the vector describing the time-dependent separation between two fluid particles is $\bm{r}(t)$, then if $\bm{r}(t)$ lies in the dissipation range of the turbulence, it follows from \eqref{llff} that
\begin{align}
\dot{\bm{r}}(t)=\bm{\Gamma}(t)\bm{\cdot}\bm{r}(t),\label{rdot}
\end{align}
where $\bm{\Gamma}(t)\equiv \bm{\Gamma}(\bm{x}(t),t)$, and $\bm{x}(t),\bm{x}(t)+\bm{r}(t)$ denote the positions of the two particles. We are concerned with incompressible flows where $\mathrm{tr}[\bm{\Gamma}(t)]=0$. We shall also focus on statistically stationary, isotropic turbulence.

Defining $r(t)\equiv \|\bm{r}(t)\|$, then from \eqref{rdot} we obtain
\begin{align}
\dot{{r}}(t)=\mathcal{S}(t)r(t),\label{rdot2}
\end{align}
where $\mathcal{S}(t)\equiv(\bm{e}(t)\bm{\cdot}\bm{\Gamma}(t))\bm{\cdot}\bm{e}(t)\equiv (\bm{e}(t)\bm{\cdot}\mathbf{S}(t))\bm{\cdot}\bm{e}(t)$, $\mathbf{S}\equiv (1/2)(\bm{\Gamma}+\bm{\Gamma}^\top)$ is the strain-rate tensor, and $\bm{e}(t)\equiv \bm{r}(t)/r(t)$. 

The classical argument (e.g. \cite{jullien03,rivera05,salazar09}) for an exponential separation of fluid particle-pairs in the dissipation range is that since according to \eqref{rdot2}, $\dot{{r}}(t)\propto r(t)$, then $r(t)$ must grow exponentially. If the effect of $\mathcal{S}(t)$ is described phenomenologically, then one obtains the following expression for the FIT mean-square separation
\begin{align}
\Big\langle{r}^2(t)\Big\rangle_\xi=\xi^2e^{\mathcal{A}^F t/\tau_\eta},\label{mssES}
\end{align}
where $\mathcal{A}^F>0$ is a non-dimensional constant that captures the relevant properties of $\mathcal{S}$, and $\tau_\eta$ is the Kolmogorov timescale. We will not discuss at this point the specification of $\mathcal{A}^F$ since here we are mainly interested in the question of the validity of the functional form of the prediction in \eqref{mssES}. The same kind of phenomenology would also imply that \eqref{mssES} applies to the BIT case $\langle{r}^2(-t)\rangle_\xi$, except with  $\mathcal{A}^F$ replaced by the BIT constant $\mathcal{A}^B$ giving $\langle{r}^2(-t)\rangle_\xi=\xi^2e^{\mathcal{A}^B t/\tau_\eta}$, and one would expect $\mathcal{A}^B>\mathcal{A}^F$ for 3D turbulence \cite{buaria15,bragg16}.

This kind of phenomenological approach to deriving predictions for $\langle{r}^2(t)\rangle_\xi$ and  $\langle{r}^2(-t)\rangle_\xi$ may fail, however, since $\mathcal{S}(t)$ is not a constant in turbulence, and hence the form of the solution to \eqref{rdot2} may not be that of an exponential with an exponent linear in $t$. This can be seen more clearly by considering the solution to \eqref{rdot2} 
\begin{align}
{r}(t)=r(0)\exp\Bigg(\int_0^t\mathcal{S}(t')\,dt'\Bigg),\label{rdotsol}
\end{align}
from which we obtain expressions for the $N^{th}$ moment of the FIT and BIT separation
\begin{align}
\Big\langle{r}^N(t)\Big\rangle_\xi&=\xi^N\Bigg\langle\exp\Bigg(N\int_0^t\mathcal{S}(t')\,dt'\Bigg)\Bigg\rangle_\xi,\label{mss}\\
\Big\langle{r}^N(-t)\Big\rangle_\xi&=\xi^N\Bigg\langle\exp\Bigg(-N\int^0_{-t}\mathcal{S}(t')\,dt'\Bigg)\Bigg\rangle_\xi.\label{mssB}
\end{align}
In principle, the term $\langle\exp(N\int_0^t\mathcal{S}(t')\,dt' )\rangle_\xi$ only reduces to a form such as $e^{\mathcal{A}^Ft/\tau_\eta}$ if $\mathcal{S}$ is a constant (and similarly for the BIT case), which it is not in turbulence, nor can it be approximated as being so over the observation time of the pair-dispersion. Nevertheless, it is often stated in the literature, without qualification, that in the dissipation range $\langle{r}^2(t)\rangle_\xi= \xi^2 e^{\mathcal{A}^Ft/\tau_\eta}$, e.g. \cite{salazar09,rivera05,ni13,costa17}. In other works, it is stated that $\langle{r}^2(t)\rangle_\xi= \xi^2 e^{\mathcal{A}^F t/\tau_\eta}$ is the form only in the long-time regime $t\gg\tau_\eta$, e.g. \cite{sawford01,falkovich01,borgas04}. Clearly then, there is some disagreement and confusion in the literature over the validity of \eqref{mssES}. This is undoubtedly in part because phenomenological approaches to deriving $\langle{r}^2(t)\rangle_\xi= \xi^2 e^{\mathcal{A}^Ft/\tau_\eta}$, such as discussed earlier, do not distinguish the regime of $t$ for which this should be valid, whereas more formal methods, such as those discussed in \cite{falkovich01}, only lead to $\langle{r}^2(t)\rangle_\xi= \xi^2 e^{\mathcal{A}^Ft/\tau_\eta}$ in the long-time regime $t\gg\tau_\eta$.

In view of these issues, we now consider in detail the behavior of $\langle{r}^N(t)\rangle_\xi$ and $\langle{r}^N(-t)\rangle_\xi$, first in the short-time regime, and then in the long-time regime.
\subsection{Dispersion in the short-time regime}\label{STR}
The exact solution in \eqref{mss} can be expanded to give 
\begin{align}
\begin{split}
\Big\langle{r}^N(t)\Big\rangle_\xi=\xi^N\Bigg(1&+N\int_0^t\Big\langle\mathcal{S}(t')\Big\rangle_\xi\,dt'+\frac{N^2}{2}\int_0^t\int_0^t\Big\langle\mathcal{S}(t')\mathcal{S}(t'')\Big\rangle_\xi\,dt'\,dt''\\
&+\frac{N^3}{6}\int_0^t\int_0^t\int_0^t\Big\langle\mathcal{S}(t')\mathcal{S}(t'')\mathcal{S}(t''')\Big\rangle_\xi\,dt'\,dt''dt'''+\cdot\cdot\cdot\Bigg).
\end{split}
\label{mssST}
\end{align}
For $N=2$, there is no reason to expect $\langle{r}^2(t)\rangle_\xi= \xi^2 e^{\mathcal{A}^F t/\tau_\eta}$ to be true for arbitrary $t$ for the simple reason that the terms on the rhs of \eqref{mssST} are not in general related to each other as they would be for an exponential function of the form $e^{\mathcal{A}^Ft/\tau_\eta}$. For example, there is no reason to expect that
\begin{align}
\Bigg(\int_0^t\Big\langle\mathcal{S}(t')\Big\rangle_\xi\,dt'\Bigg)^2=\int_0^t\int_0^t\Big\langle\mathcal{S}(t')\mathcal{S}(t'')\Big\rangle_\xi\,dt'\,dt'',
\end{align}
unless either $\mathcal{S}$ is a constant (which it is certainly not in turbulence), or else if $\int_0^t\mathcal{S}(t')\,dt'$ is non-random. The latter is in general not true for finite $t$. These considerations would then rule out the possibility that $\langle{r}^2(t)\rangle_\xi= \xi^2 e^{\mathcal{A}^Ft/\tau_\eta}$ is correct for arbitrary $t$.

The leading order behavior of the rhs of \eqref{mssST} comes from using the leading-order contribution to $\mathcal{S}(t)$ in the limit $t\to 0$. To be fully consistent in the order of the approximation of the resulting integrals we must take $\mathcal{S}(t)=\mathcal{S}(0)+t\dot{\mathcal{S}}(0)+\mathcal{O}(t^2)$ and then obtain

\begin{align}
\Big\langle{r}^N(t)\Big\rangle_\xi=\xi^N\Bigg(1+Nt\Big\langle\mathcal{S}(0)\Big\rangle_\xi+\frac{N}{2} t^2\Big\langle\dot{\mathcal{S}}(0)\Big\rangle_\xi    +\frac{N^2}{2}t^2\Big\langle\mathcal{S}(0)\mathcal{S}(0)\Big\rangle_\xi +\mathcal{O}(t^3)\Bigg),\quad\text{for}\, t\to 0.\label{mssSTb}
\end{align}
Since the strain-rate $\mathcal{S}(t)$ has a time scale $\mathcal{O}(\tau_\eta)$, the expansion in \eqref{mssSTb} should be strictly understood as an expansion in $t/\tau_\eta$.

Our concern is first with $\langle\mathcal{S}(0)\rangle_\xi$ since it is the presence of the term in \eqref{mssSTb} involving this that would dominate the growth of $\langle{r}^N(t)\rangle_\xi$ in the limit $t\to 0$ if $\langle\mathcal{S}(0)\rangle_\xi\neq0$.

A correct handling of the conditional average $\langle\mathcal{S}(0)\rangle_\xi$ requires careful consideration of the state of the system at time $t=0$, where we remind the reader that $t=0$ does not necessarily correspond to the initial time of the whole flow, but rather denotes the time at which the fluid particles are ``marked''. There are two cases of interest; 1) where the particles are already in the system at $t<0$ and then one simply ``marks'' particle-pairs that satisfy $r(0)=\xi$, and records their separation at $t>0$, and 2) where the particles are introduced to the system at time $t=0$ with the specified initial separation $r(0)=\xi$. While the latter is usually the case considered in pair-dispersion studies, the former is of interest when one wishes to consider how particles that are already in the flow are mixing, i.e. it provides insight into the turbulent flow itself through the Lagrangian perspective.

We now consider the first case. In incompressible turbulence, if tracer particles are introduced to the flow at some time $t<0$ with a non-uniform spatial distribution, then eventually their spatial distribution will become uniform and constant in time, and will remain so due to incompressibility of the flow \cite{bragg12b}. If at time $t=0$ the particles have attained this ``fully-mixed'' asymptotic regime, then $\delta (r(0)-\xi)$ is not random, such that $\langle \mathcal{S}(0)\delta (r(0)-\xi)\rangle=\langle\mathcal{S}(0)\rangle \delta (r(0)-\xi)$, and
\begin{align}
\Big\langle \mathcal{S}(0)\Big\rangle_\xi=\Big\langle \mathcal{S}(0)\Big\rangle\equiv \Big\langle(\bm{e}(0)\bm{\cdot}\bm{\Gamma}(0))\bm{\cdot}\bm{e}(0)\Big\rangle.\label{FMav}
\end{align}
This average may not be zero in isotropic turbulence under the conditions we are here considering since $\bm{e}(0)\equiv r^{-1}(0)\bm{r}(0)$ depends upon $\bm{\Gamma}(t<0)$, which is correlated with $\bm{\Gamma}(0)$ in turbulence. Put another way, the average in \eqref{FMav} may not be zero because the separation direction of the particle-pair is typically correlated with the instantaneous eigenframe of $\bm{\Gamma}$.

In the case where the tracer particles are not fully-mixed at time $t=0$, $\delta (r(0)-\xi)$ is random, so that in general $\langle\mathcal{S}(0)\rangle_\xi\neq\langle\mathcal{S}(0)\rangle$, and $\langle\mathcal{S}(0)\rangle_\xi$ is non-zero \cite{swailes97,swailes99}. These considerations therefore show that in the case where the particle-pairs are marked at $t=0$, rather than introduced at $t=0$ with a specified initial separation, the leading order term that determines the growth of $\langle{r}^N(t)\rangle_\xi$ in the limit $t\to 0$ is $\propto t$.

%If we assume the particles are introduced such that there is no correlation between $r(0)$ and $\mathcal{S}(0)$ then the conditional averages reduce to unconditioned averages.

If we consider fluid particle-pairs that are introduced with the specified initial separation $r(0)=\xi$, then $\delta (r(0)-\xi)$ is not random, and so just as for the ``fully-mixed'' case,  $\langle \mathcal{S}(0)\delta (r(0)-\xi)\rangle=\langle\mathcal{S}(0)\rangle \delta (r(0)-\xi)$ and \eqref{FMav} applies. If there is no correlation between the initial pair-separation direction and $\bm{\Gamma}(0)$, then for isotropic turbulence we obtain
\begin{align}
\Big\langle\mathcal{S}(0)\Big\rangle\equiv \Big\langle(\bm{e}(0)\bm{\cdot}\bm{\Gamma}(0))\bm{\cdot}\bm{e}(0)\Big\rangle= \Big\langle\bm{e}(0)\bm{e}(0)\Big\rangle \bm{:}\Big\langle \bm{\Gamma}(0)\Big\rangle={0}.
\end{align}
On the other hand, if the particle-pairs are introduced to the system with some correlation between the initial pair-separation direction and $\bm{\Gamma}(0)$ then $\langle\mathcal{S}(0)\rangle\neq 0$.

Together with \eqref{mssSTb}, these observations imply that \emph{if} there is no correlation between the initial pair-separation direction $\bm{e}(0)$ and $\bm{\Gamma}(0)$, then 
\begin{align}
\Big\langle{r}^N(t)\Big\rangle_\xi -\xi^N\propto t^2,\quad\text{for}\, t\to 0,\label{mssT2}
\end{align}
but otherwise
\begin{align}
\Big\langle{r}^N(t)\Big\rangle_\xi -\xi^N\propto t,\quad\text{for}\, t\to 0.\label{mssT}
\end{align}
The important point then is that the leading order behavior of $\langle{r}^N(t)\rangle_\xi$ depends crucially upon the statistical state of the system at time $t=0$. In many situations, however, it might be difficult to observe the $\propto t$ contribution unless  $\langle\mathcal{S}(0)\rangle$ is sufficiently large. In \S\ref{RD} we will perform a test case to consider this.

%The behavior $\langle{r}^2(t)\rangle_\xi= \xi^2 e^{\mathcal{A}^F t/\tau_\eta}$ is trivially consistent with the behavior in \eqref{mssT} for $N=2$ in the limit $t\to0$, however, this is hardly grounds for claiming that $\langle{r}^2(t)\rangle_\xi= \xi^2 e^{\mathcal{A}^F t/\tau_\eta}$ is correct in the short-time limit.

Concerning the BIT case $\langle{r}^N(-t)\rangle_\xi$, the same arguments apply, except with one important difference. If $\langle\mathcal{S}(0)\rangle_\xi=0$, then $\langle{r}^N(-t)\rangle_\xi -\langle{r}^N(t)\rangle_\xi\propto t^3$ in the limit $t\to 0$, but if $\langle\mathcal{S}(0)\rangle_\xi\neq0$, then $\langle{r}^N(-t)\rangle_\xi -\langle{r}^N(t)\rangle_\xi\propto t$ in the limit $t\to 0$.
%%
%\subsection{Intermediate-times}
%%
%The result \eqref{mssSTe} shows that $\langle{r}^2(t)\rangle_\xi\approx \xi^2 e^{\mathcal{A}t/\tau_\eta}$ could be considered asymptotically correct in the limit $t\to0$ \emph{if} the initial particle separations are correlated with the eigenframe of $\bm{\Gamma}(0)$. However, at finite times, the contributions from higher-order terms would render $\langle{r}^2(t)\rangle_\xi\approx \xi^2 e^{\mathcal{A}t/\tau_\eta}$ and invalid description. This can be seen by considering \eqref{mssST}; since in general
%%
%\begin{align}
%2\Bigg(\int_0^t\Big\langle\mathcal{S}(t')\Big\rangle_\xi\,dt'\Bigg)^2\not\propto 2\int_0^t\int_0^t\Big\langle\mathcal{S}(t')\mathcal{S}(t'')\Big\rangle_\xi\,dt'\,dt'',
%\end{align}
%%
%and similarly for the higher-order terms in \eqref{mssST}, then the series in parenthesis on the rhs of \eqref{mssST} cannot in general be re-summed to give an exponential function of the form $e^{\mathcal{A}t/\tau_\eta}$.
%
\subsection{Dispersion in the long-time regime}\label{LTR}
Batchelor \cite{batchelor52a} was the first to derive a prediction for the mean separation $\langle{r}(t)\rangle_\xi$ in the regime $t\gg\tau_\eta$. He argued that for $t\gg\tau_\eta$, the effect of the initial direction vector $\bm{e}(0)$ would be lost, and $\mathcal{S}(t)$ would become a stationary random function of $t$, with time-averaged mean value $\mu^F\equiv \lim_{t\to\infty}[t^{-1}\int_0^t\mathcal{S}(t')\,dt']$ (assumed positive), leading to
\begin{align}
\Big\langle{r}(t)\Big\rangle_\xi=\xi e^{\mu^F  t},\quad\text{for}\, t\gg\tau_\eta.\label{Bmss}
\end{align}
Batchelor went on to argue more generally that the order-of-magnitude estimate for $r(t)$ in the regime $t\gg\tau_\eta$ is $r(t)\sim r(0)  e^{\mu^F  t}$, from which we obtain
\begin{align}
\Big\langle{r}^N(t)\Big\rangle_\xi\sim\xi^N e^{N\mu^F  t},\quad\text{for}\, t\gg\tau_\eta.\label{NsepB}
\end{align}
Batchelor pointed out that it is possible for the estimate $r(t)\sim r(0)  e^{\mu^F  t}$ (and therefore \eqref{NsepB}) to fail, since it is always possible that fluctuations of $t^{-1}\int_0^t\mathcal{S}(t')\,dt'$ about $\mu^F$ could be significant in certain realizations of the system, e.g. during extreme events in the turbulence \cite{scatamacchia12}.

The effect of fluctuations in $\int_0^t\mathcal{S}(t')\,dt'$ can be accounted for in the long-time regime by appealing to the Central Limit Theorem (CLT) \cite{borgas04,balkovsky99,falkovich01}. In the regime $t\gg \tau_\eta$, one may argue that the integral $\int_0^t\mathcal{S}(t')\,dt'$ behaves as the sum of a large number, $\mathcal{O}(t/\tau_\eta)$, of independent, identically distributed random numbers, with mean value $\mu^F t$. Under these conditions the CLT may be invoked \cite{borgas04}, according to which the fluctuations of $\int_0^t\mathcal{S}(t')\,dt'$ about $\mu^F t$ are normally distributed with variance $\beta\Sigma_\mathcal{S} \tau_\eta t$, where $\beta=\mathcal{O}(1)$, and $\Sigma_\mathcal{S}$ is the variance of $\mathcal{S}$. By integrating over the normal distribution of these fluctuations, the following may be derived \cite{buaria15}
	\begin{align}
	\Big\langle{r}^N(t)\Big\rangle_\xi&=\xi^N e^{\zeta_Nt},\quad\text{for}\, t\gg\tau_\eta,\label{Nsep}\\
	\zeta_N&\equiv N\mu^F   +(N^2/2) \beta\Sigma_\mathcal{S} \tau_\eta.
	\end{align}
The result in \eqref{Nsep} has the same functional form as \eqref{NsepB}, and as expected, is identical to \eqref{NsepB} in the absence of fluctuations of $\int_0^t\mathcal{S}(t')\,dt'$ about $\mu^F t$, i.e. when $\Sigma_\mathcal{S}=0$. As pointed out in \cite{buaria15}, using the CLT to derive $\langle{r}^N(t)\rangle_\xi$ for $t\gg \tau_\eta$ is not exact, as it ignores extreme events in the behavior of $\mathcal{S}$. Due to the intermittent nature of turbulence \cite{frisch}, deviations from \eqref{Nsep} are possible and could be important, especially for larger $N$ and/or as $R_\lambda$ increases. To capture the effect of such extreme fluctuations, one would need to use large deviation theory \cite{falkovich01}.

The results discussed so far are effectively for the regime $t\gg\tau_\eta$, where fluctuations of $\int_0^t\mathcal{S}(t')\,dt'$ about the mean value can play an important role. However, an important question is how $\int_0^t\mathcal{S}(t')\,dt'$ behaves asymptotically as $t\to\infty$. By definition, as $t\to\infty$, $\int_0^t\mathcal{S}(t')\,dt'\to\mu^F t$. In the context of dynamical systems theory, the quantity $\mu^F$ is referred to as the Lyapunov exponent \cite{ott02}, and equivalent to the definition $\mu^F\equiv \lim_{t\to\infty}[t^{-1}\int_0^t\mathcal{S}(t')\,dt']$ is
\begin{align}
\mu^F\equiv\lim_{t\to \infty}\lim_{r(0)\to 0}\frac{1}{t}\ln\Bigg(\frac{r(t)}{r(0)}\Bigg).
\end{align}
Since $\mu^F$ is defined for a single realization of $r(t)$, it may in principle vary for differing realizations of the system. From this it follows that even though in a given realization the asymptotic growth is exponential $r(t)= r(0)  e^{\mu^F  t}$ as $t\to\infty$, the moments $\langle r^N(t)\rangle_\xi$ need not be since $\langle e^{N\mu^F t}\rangle_\xi \neq e^{Nt \langle \mu^F \rangle_\xi} $ in general. However, if $\mathcal{S}$ is ergodic then $\mu^F$ is the same for every realization of the system \cite{falkovich01}, and it follows exactly that for $t\to\infty$, $\langle{r}^N(t)\rangle_\xi =\xi^N e^{N\mu^F  t}$. Therefore, failure to observe $\langle{r}^N(t)\rangle_\xi = \xi^N e^{N\mu^F  t}$ in a turbulent flow can only be either because $t$ is not sufficiently large, or else $\mathcal{S}(t)$ is not-ergodic.
 Although there exists no formal proof that turbulent flows are ergodic, it is a standard to assume that they are \cite{monin1}. The numerical study in \cite{galanti04} presented evidence for ergodicity in stationary, homogeneous turbulence, but showed that very long times are required for this to be observed. Furthermore, from a pragmatic perspective, even if ergodicity is satisfied in homogeneous turbulence, the integral $(1/t)\int_0^t\mathcal{S}(t')\,dt'$ may converge to a realization-independent value very slowly with increasing $t$ due to the small-scale intermittency of turbulence.
 
Summarizing then, in the regime $t\gg \tau_\eta$, application of the CLT predicts $\langle{r}^N(t)\rangle_\xi \approx\xi^N e^{\zeta_N  t}$, which may however fail due to the neglect of large deviations in $\int_0^t\mathcal{S}(t')\,dt'$ about $\mu^F t$. However, in the asymptotic limit $t\to\infty$, $\langle{r}^N(t)\rangle_\xi =\xi^N e^{N\mu^F  t}$ must of necessity be recovered if $\mathcal{S}$ is ergodic.

For the BIT case, the same arguments apply only that now in place of $\mu^F$ we have $\mu^B\equiv \lim_{t\to\infty}[t^{-1}\int^0_{-t}\mathcal{S}(t')\,dt']$. Whereas $\mu^F>0$, we expect that $\mu^B<0$ so that $\langle{r}^N(-t)\rangle_\xi =\xi^N e^{-N\mu^B  t}$ grows with increasing $t$. Furthermore, in 3D turbulence where it is known that BIT dispersion is faster than FIT dispersion \cite{buaria15,bragg16}, we expect that $|\mu^B|>|\mu^F|$. This then leads to an interesting behavior of the irreversibility of the dispersion, namely
\begin{align}
\Big\langle{r}^N(-t)\Big\rangle_\xi\Bigg/\Big\langle{r}^N(t)\Big\rangle_\xi&=e^{-N(\mu^F+\mu^B)t},\quad\text{for}\, t\to\infty.
\end{align}
This suggests that in 3D turbulence where $\mu^F+\mu^B<0$, the irreversibility of the relative dispersion increases exponentially as $t$ increases, leading to enormous differences between BIT and FIT dispersion. This is distinctly different from the case where the particles are in the inertial range where, if Richardson's phenomenology is correct, $\langle{r}^N(-t)\rangle_\xi /\langle{r}^N(t)\rangle_\xi$ is indepdendent of time (e.g. $\langle{r}^2(-t)\rangle_\xi /\langle{r}^2(t)\rangle_\xi=\mathfrak{g}^B /\mathfrak{g}^F$). According to our arguments in \cite{bragg16,bragg17b,bragg17c}, the irreversibility of fluid particle relative dispersion arises, fundamentally, both because $\Delta\bm{u}$ has finite temporal correlations and because the Probability Density Function (PDF) of $\Delta\bm{u}$ is asymmetric at sub-integral scales in turbulence. However, these results show that it is not only the degree of asymmetry of the PDF of $\Delta\bm{u}$ that governs how strongly irreversibile the dispersion is, but also whether the field $\Delta\bm{u}$ is smooth or rough.

\section{Evidence for exponential separation}\label{Evid}
Having considered the theoretical basis for the exponential growth of $\langle{r}^N(t)\rangle_\xi$ and $\langle{r}^N(-t)\rangle_\xi$, we now turn to consider evidence for this prediction for fluid particles dispersing in the dissipation range of turbulence. Since almost all of the purported evidence is for the FIT case, we shall mainly focus on that. 

There are a handful of claims in the literature to observe $\langle{r}^2(t)\rangle_\xi =\xi^2 e^{2\mu^F  t}$, however, several of these claims are problematic. Indeed, a recent paper \cite{afik17} presented a brief literature review of these claims, on the basis of which they conclude that there is no convincing evidence that $\langle{r}^2(t)\rangle_\xi$ grows exponentially in the dissipation range. We now discuss some of these claims in more detail, along with some additional references.

In the study of \cite{ni13}, the authors claimed to observe the behavior $\langle{r}^2(t)\rangle_\xi =\xi^2 e^{2\mu^F  t}$ in their experiments on fluid particle-pair dispersion in the dissipation range of convecting turbulent flows. However, as we previously discussed in \cite{bragg16}, their results are highly problematic since they claim to observe $\langle{r}^2(t)\rangle_\xi =\xi^2 e^{2\mu^F  t}$ in the short-time regime, for $t\leq\mathcal{O}(\tau_\eta)$, whereas as discussed in \S\ref{PPD}, such exponential growth should only occur in the long-time regime $t\gg\tau_\eta$. 

In \cite{buaria15}, Direct Numerical Simulations (DNS) were used to investigate fluid particle-pair dispersion in turbulence. They showed results for $\langle{r}^N(t)\rangle_\xi/\langle{r}^2(t)\rangle^{N/2}_\xi$ with $N=3,4$. For $\xi=\eta$, their results show behavior that is consistent with the exponential growth of $\langle{r}^N(t)\rangle_\xi$, however, the agreement is only over a transitory period, and it is difficult to tell from the plots how good the agreement is. In \cite{bragg16} we also used our DNS data for $\langle{r}^2(t)\rangle_\xi$, and plotted the quantity $t^{-1}\ln[\xi^{-2}\langle{r}^2(t)\rangle_\xi]$, which would be equal to the constant value $2\mu^F$ if the prediction $\langle{r}^2(t)\rangle_\xi =\xi^2 e^{2\mu^F  t}$  were correct. In our data we considered $\xi/\eta\in[1/4,1]$ and did not observe any regime of $t$ in which $t^{-1}\ln[\xi^{-2}\langle{r}^2(t)\rangle_\xi]$ was even approximately constant. 

There have also been claims to observe exponential growth of $\langle{r}^2(t)\rangle_\xi$ in the direct cascade regime of 2D turbulence, where the velocity field is smooth and \eqref{rdot} applies. In this context, the prediction of Lin \cite{lin72} is thought to apply, who used dimensional analysis to predict that in the direct cascade regime of 2D turbulence $\langle{r}^2(t)\rangle_\xi= \xi^2 e^{2\mathrm{A}\epsilon_\omega^{1/3}t}$, where $\mathrm{A}$ is an $\mathcal{O}(1)$ constant, and  $\epsilon_\omega$ is the mean enstrophy flux. In \cite{jullien03}, the relative dispersion of fluid particle pairs in the direct cascade regime of 2D turbulence was examined by first obtaining experimental data for the fluid velocity field at fixed grid points, and then numerically simulating fluid particle trajectories in the flow using this data. The study in \cite{jullien03} claims to observe $\langle{r}^2(t)\rangle_\xi= \xi^2 e^{2\mathrm{A}\epsilon_\omega^{1/3}t}$ in the initial stage of the dispersion, followed by a transition to a regime where $\langle{r}^2(t)\rangle_\xi$ grows as a power-law in time. However, this claim is problematic for at least two reasons. First, if one looks closely at the inset of Fig.~11 in \cite{jullien03}, it is clear that their data for $\langle{r}^2(t)\rangle_\xi$ does not exactly follow the form $\xi^2 e^{2\mathrm{A}\epsilon_\omega^{1/3}t}$. At very short times, their data grows slower than $\xi^2 e^{2\mathrm{A}\epsilon_\omega^{1/3}t}$. Beyond this, their data seems to follow $\langle{r}^2(t)\rangle_\xi= \xi^2 e^{2\mathrm{A}\epsilon_\omega^{1/3}t}$ more closely, yet only for a relatively short time, and even then their data shows a growth of $\langle{r}^2(t)\rangle_\xi$ with time that is somewhat slower than the exponential prediction. Second, the initial separations of the particles in their results are an order of magnitude smaller than the spacing of the grid points on which their experimental data for the fluid velocity field was recorded. Thus, their results could be strongly affected by interpolation errors, and furthermore, the details of the interpolation method used are not given, so that we do not know the accuracy of the method they employed. The validity of the Lin exponential prediction $\langle{r}^2(t)\rangle_\xi= \xi^2 e^{2\mathrm{A}\epsilon_\omega^{1/3}t}$ was also examined using DNS of 2D turbulence in \cite{babiano90}. In contrast to the findings of \cite{jullien03}, the study in \cite{babiano90} states that the Lin exponential prediction is, strictly speaking, never observed in their DNS data.

In \cite{rivera05}, an experimental setup similar to that used in \cite{jullien03} was employed to study pair dispersion in 2D turbulence. Their results for $\langle{r}^2(t)\rangle_\xi$ with $\xi$ in the direct cascade regime did not show evidence of the expected exponential growth, however, they argued that this was due to scale contamination. That is, $\langle{r}^2(t)\rangle_\xi$ may involve contributions from particle-pairs that are not only in the direct cascade regime at time $t$, but also some that have left this regime, and this mixing of different scales contaminates the results. In an attempt to avoid this problem, they instead used doubling-time statistics to look for evidence of the exponential separation of particle-pairs in the direct cascade regime. In the doubling-time method, instead of analyzing pair-dispersion through quantities such as $\langle{r}^2(t)\rangle_\xi$, one computes the time $T_\rho$ that it takes for particle-pairs with separation $\xi $ to reach the separation $\rho \xi$. With this method it is then possible to measure the dispersion characteristics of particle-pairs confined a desired range of scales. For $r(t)$ in the smooth-regime of the fluid velocity field we may express $T_\rho$ through the integral
\begin{align}
\ln[\rho]=\int_0^{T_\rho}\mathcal{S}(t')\,dt',\label{DT}
\end{align}
such that $T_\rho$ is a random variable that depends upon the realization of $\mathcal{S}$. In \cite{rivera05} the quantity $\langle T_\rho\rangle$ was measured using $\rho=1.2$ and they found that it was nearly constant for $\xi$ below the energy injection scale $r_{inj}$ of the flow (where the velocity field is smooth), with the value $\langle T_\rho\rangle=1.04\tau_\omega$, where $\tau_\omega\equiv\sqrt{ \langle\|\bm{\omega}\|^2\rangle}$, and ${\bm{\omega}\equiv \bm{\nabla}\times\bm{u}}$ is the vorticity. They concluded that the independence of $\langle T_\rho\rangle$ for $\xi<r_{inj}$ indicates exponential separation of the fluid particles. However, it is not clear to us that this conclusion follows. In particular, as can be seen from \eqref{DT}, the constancy of $\langle T_\rho\rangle$ for $\xi<r_{inj}$ is simply because in this smooth regime of the fluid velocity field, $T_\rho$ is governed by $\mathcal{S}$ which is scale-independent. Put another way, constancy of $\langle T_\rho\rangle$ for $\xi<r_{inj}$ implies the expected result $\dot{r}(t)\propto r(t)$, but this does not imply exponential growth of $r(t)$ since the proportionality variable (i.e. $\mathcal{S}$) fluctuates in time. The authors of \cite{rivera05} also argue that the fact that $\langle T_\rho\rangle\approx \tau_\omega$ strongly indicates exponential growth of the pair-separation for $\xi<r_{inj}$, noting that for homogeneous turbulence we also have $\tau_\omega= \sqrt{ \langle\|\bm{S}\|^2\rangle}$. However, this again does not imply that $r(t)$ grows exponentially in time, but only that $\langle T_\rho\rangle$ is dominated by the characteristic properties of the straining field.

In agreement with \cite{afik17}, we therefore conclude that there is no convincing evidence for the validity of $\langle{r}^2(t)\rangle_\xi =\xi^2 e^{2\mu^F  t}$, despite the fact that this result is often cited in the literature as essentially fact (e.g. \cite{sawford01,salazar09,ni13,costa17}).

In closing this section, we note that previous studies such as \cite{gir90b,goto07} computed the quantity $\langle (d/dt)\ln[r(t)]\rangle_\xi\equiv \langle \mathcal{S}(t)\rangle_\xi$ using DNS of isotropic turbulence, and found that $\langle (d/dt)\ln[r(t)]\rangle_\xi$ became constant for times greater than a few multiples of $\tau_\eta$. This then indicates that $r(t)$ does grow exponentially in time after the short-time regime. However, it does not follow from this that $\langle{r}^N(t)\rangle_\xi$ grows exponentially, since $\langle (d/dt)\ln[r(t)]\rangle_\xi$ is in a sense only a low-order measure of the dispersion. For example, whereas by definition, $\langle (d/dt)\ln[r(t)]\rangle_\xi$ only depends upon the mean of $\mathcal{S}(t)$, $\langle{r}^N(t)\rangle_\xi$ is also affected by the fluctuations of $\mathcal{S}(t)$ about this mean through the integral $\int_0^t\mathcal{S}(t')\,dt'$. It therefore remains that there exists no convincing evidence that $\langle{r}^N(t)\rangle_\xi$ grows exponentially in the dissipation range of turbulence (and similarly for the BIT case).
\section{Test for exponential separation}\label{test}
As just summarized in \S\ref{Evid}, to the best of our knowledge, $\langle{r}^N(t)\rangle_\xi =\xi^N e^{N\mu^F  t}$ (and its BIT counterpart) has never convincingly been observed in turbulence, even for $N=2$. One of the difficulties in testing $\langle{r}^N(t)\rangle_\xi =\xi^N e^{N\mu^F  t}$ is that if $r(0)/\eta$ is finite, then in the regime $t\gg\tau_\eta$ the particle-pairs will have long since left the dissipation regime, so that the linear equation \eqref{rdot}, upon the basis of which $\langle{r}^N(t)\rangle_\xi =\xi^N e^{N\mu^F  t}$ is derived, is no longer valid. However, an implication of \eqref{rdotsol} is that since
\begin{align}
\frac{{r}(t)}{\eta}=\frac{r(0)}{\eta}\exp\Bigg(\int_0^t\mathcal{S}(t')\,dt'\Bigg),\label{rdotsol2}
\end{align}
then we can satisfy $r(t)\ll\eta$ even for $t\gg\tau_\eta$ simply by making $r(0)/\eta$ small enough. More formally, \eqref{rdot} remains valid for $t\in[0,\infty)$ in the limit ${r}(0)\to 0$, as was also mentioned in \cite{batchelor52a}. Consequently, if the long-time result $\langle{r}^N(t)\rangle_\xi =\xi^N e^{N\mu^F  t}$ is correct, then it should always be observable in a turbulent flow provided $r(0)/\eta$ is sufficiently small. However, using $r(0)/\eta\lll 1$ in experiments and DNS is problematic. In experiments, the minimum $r(0)/\eta$ that can be considered is restricted, among other factors, by the finite size of the tracer particles used. In DNS, pair-dispersion is usually examined by solving the trajectories of the individual fluid particles, and then subtracting their positions to calculate $r(t)$. However, if their separation is much smaller than the grid resolution in the DNS, then the calculated relative motion of the particle-pair could be strongly affected by the interpolation methods used in DNS to interpolate the fluid velocities from the grid points to the particle positions. Indeed, the interpolation errors could violate the linearity of the local flow (i.e. fail to preserve $\dot{r}(t)\propto r(t)$) that the particles should experience when $r(t)\ll\eta$.

To circumvent these problems, we propose that instead of using DNS to simulate the trajectories of individual particles, and then subtracting their positions to compute $\bm{r}(t)$, we may instead directly solve \eqref{rdot}. In this way, the turbulence is specified through the single-point quantity $\bm{\Gamma}$, and as discussed by Ray \& Collins \cite{ray13}, such a method eliminates the aforementioned interpolation errors which could otherwise strongly affect the relative motion results when the particle separations are much smaller than the DNS grid spacing. 

We therefore solve \eqref{rdot} using DNS data to prescribe $\bm{\Gamma}(t)$, and we do this for $262 144$ realizations of $\bm{\Gamma}(t)$, from which we compute $\langle{r}^N(t)\rangle_\xi$ and other relevant statistics.  In addition, we also compute the BIT pair-dispersion statistics $\langle{r}^N(-t)\rangle_\xi$ that are obtained by solving the time-reversed form of \eqref{rdot}. Our DNS is for statistically stationary, isotropic turbulence generated in a periodic box of length $2\pi$, and since we wish to consider the dispersion up to very long times we consider $R_\lambda=90$, solved on a grid of size $128^3$. Details of the DNS and the numerical methodologies can be found in \cite{ireland13,ireland16a}.

When solving \eqref{rdot} we must prescribe $\bm{r}(0)$, or equivalently $r(0)$ and $\bm{e}(0)$. Although ${r}(0)\to 0$ is required to formally justify the use of \eqref{rdot} for $t\in[0,\infty)$, when solving \eqref{rdot}, the actual value of ${r}(0)$ is irrelevant since the rescaled solution $r^{-1}(0)r(t)$ is independent of $r(0)$ (see \eqref{rdotsol}). As discussed in \S\ref{STR}, the short-time behavior of $\langle{r}^N(t)\rangle_\xi$ can depend upon the statistical properties of $\bm{e}(0)$, and we therefore consider two cases in order to explore the effect. First, we select $\bm{e}(0)$ independently from $\bm{\Gamma}(0)$ using a uniform random distribution on the sphere (we shall refer to this as ``Initial Condition A''). Second, we choose $\bm{e}(0)$ to align with the eigenvector of $\bm{\Gamma}(0)$ that corresponds to its maximum eigenvalue (we shall refer to this as ``Initial Condition B''). 
%As discussed earlier, in a real turbulent flow, for finite $r(0)$, the validity of \eqref{rdot} will break down for describing fluid particle-pair separation since for sufficiently large $t$, $r(t)$ will become too large for the linear flow field assumption to apply. However, in the limit $r(0)/\eta\to 0$, \eqref{rdot} would remain valid even for $t\to\infty$ \cite{sawford01}. In this way, our test case allows us to probe whether $\langle{r}^2(t)\rangle_\xi\approx \xi^2 e^{\mathcal{A}t/\tau_\eta}$ would really occur if $r(0)/\eta\to 0$, while avoiding the numerical issues that would arise if trying to track the fluid particles individually and subtracting their positions, rather than solving \eqref{rdot} directly.

%
\section{Results \& discussion}\label{RD}
\begin{figure}[ht]
	\centering
	\vspace{-30mm}
	{\begin{overpic}
			[trim = 0mm 50mm 0mm 0mm,scale=0.4,clip,tics=20]{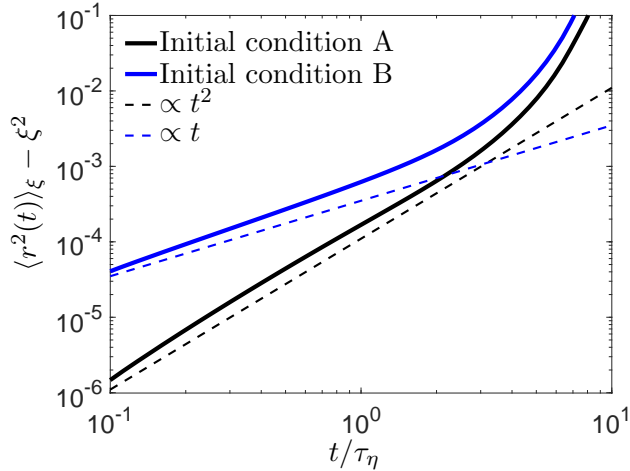}
			\put(115,8){$t/\tau_\eta$}
			\put(-5,80){\rotatebox{90}{$\langle{r}^2( t)\rangle_\xi -\xi^2$}}
			\put(51,163){\small{Initial condition A}}
			\put(51,151){\small{Initial condition B}}
			\put(51,140){\small{$\propto t^2$}}	
			\put(51,129){\small{$\propto t$}}			
	\end{overpic}}
	\caption{DNS data for $\langle{r}^2( t)\rangle_\xi -\xi^2$ using two different kinds of initial conditions, focusing on the short-time behavior. }
	\label{ballistic}
\end{figure}
\FloatBarrier
We now consider results from the simulations described in \S\ref{test}. The results in Fig.~\ref{ballistic} show that in the short-time regime, the growth of $\langle{r}^2(t)\rangle_\xi$ depends essentially upon the statistical state of the system at $t=0$. However, for both sets of results, $\langle{r}^2(t)\rangle_\xi$ definitely does not grow exponentially at short-times, in agreement with the arguments in \S\ref{STR}. These results then contradict the findings in \cite{ni13,jullien03} that claim to observe exponential growth of $\langle{r}^2(t)\rangle_\xi$ for $t/\tau_\eta\leq\mathcal{O}(1)$. 

In agreement with our analysis in \S\ref{STR}, when $\bm{e}(0)$ and $\bm{\Gamma}(0)$ are correlated, $\langle{r}^2(t)\rangle_\xi-\xi^2 \propto t$ in the limit $t\to 0$. Actually, our data shows that the growth is slightly faster than $\propto t$ down to $t/\tau_\eta =0.1$, which may be due to the contribution from the next term in the $t$ expansion of $\langle{r}^2(t)\rangle_\xi-\xi^2$ which grows as $t^2$. As explained in \S\ref{STR}, the case where $\bm{e}(0)$ and $\bm{\Gamma}(0)$ are correlated is of interest in some situations, such as when $\langle{r}^2(t)\rangle_\xi$ is used to quantify how particles that are not fully-mixed in the system at $t=0$ subsequently mix as $t$ increases. Of course, our ``Initial condition B'' is an extreme example where a strong correlation between $\bm{e}(0)$ and $\bm{\Gamma}(0)$ has been prescribed. For the situations discussed in \S\ref{STR}, the correlations would not be so strong and the duration of time for which $\langle{r}^2(t)\rangle_\xi-\xi^2 \propto t$ might be observed may be very short and difficult to observe. With Initial condition A, where $\bm{e}(0)$ and $\bm{\Gamma}(0)$ are uncorrelated, $\langle{r}^2(t)\rangle_\xi-\xi^2 \propto t^2$, i.e. ballistic growth, which is confirmed quite well in Fig.~\ref{ballistic}. As expected, the results show that the effect of the initial condition on $\langle{r}^2(t)\rangle_\xi$ disappears at long-times.
\begin{figure}[ht]
	\centering
	\vspace{-40mm}
%	\hspace{30mm}
	{\hspace{-20mm}\begin{overpic}
			[trim = 15mm 90mm 0mm 0mm,scale=1,clip,tics=20]{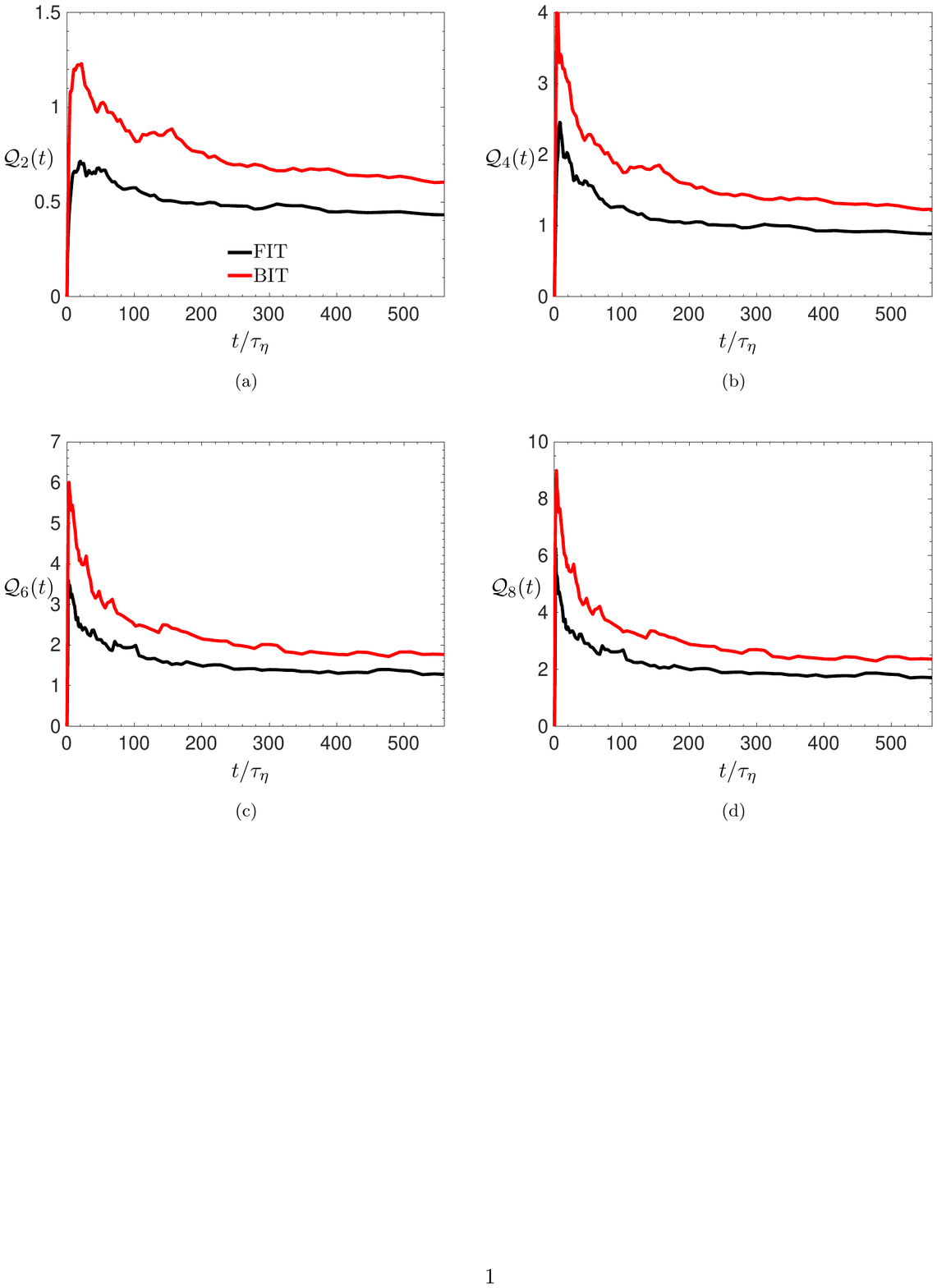}
	\end{overpic}}	
	\caption{DNS data for (a) $\mathcal{Q}_2(t)$, (b) $\mathcal{Q}_4(t)$, (c) $\mathcal{Q}_6(t)$, and $\mathcal{Q}_8(t)$, where $\mathcal{Q}_N(t)=(\tau_\eta/t)\ln[\xi^{-N}\langle{r}^N(t)\rangle_\xi]$. The BIT results correspond to $\mathcal{Q}_N(-t)=(\tau_\eta/t)\ln[\xi^{-N}\langle{r}^N(-t)\rangle_\xi]$. If $\langle{r}^N( t)\rangle_\xi= \xi^N e^{N\mu^F t}$, then $\mathcal{Q}_N(t)=N\tau_\eta\mu^F$, i.e. constant.}
	\label{r2_r4}
\end{figure}
\FloatBarrier
We now turn to consider the long-time behavior of $\langle{r}^N(t)\rangle_\xi$, both FIT and BIT. As discussed in \S\ref{LTR}, the predictions $\langle{r}^N( t)\rangle_\xi= \xi^N e^{N\mu^F t}$ and $\langle{r}^N(-t)\rangle_\xi= \xi^N e^{-N\mu^B t}$ are expected to apply at sufficiently large $t$. In Fig.~\ref{r2_r4} we plot the results in the form $\mathcal{Q}_N(t)=(\tau_\eta/t)\ln[\xi^{-N}\langle{r}^N(t)\rangle_\xi]$ and $\mathcal{Q}_N(-t)=(\tau_\eta/t)\ln[\xi^{-N}\langle{r}^N(-t)\rangle_\xi]$, such that the curves should be constant in time with $\mathcal{Q}_N(t)=N\tau_\eta\mu^F$ if $\langle{r}^N( t)\rangle_\xi= \xi^N e^{N\mu^F t}$ is correct (and similarly for the BIT case). 
\begin{figure}[ht]
	\centering
	\vspace{-30mm}
	{\begin{overpic}
			[trim = 15mm 90mm 0mm 0mm,scale=1,clip,tics=20]{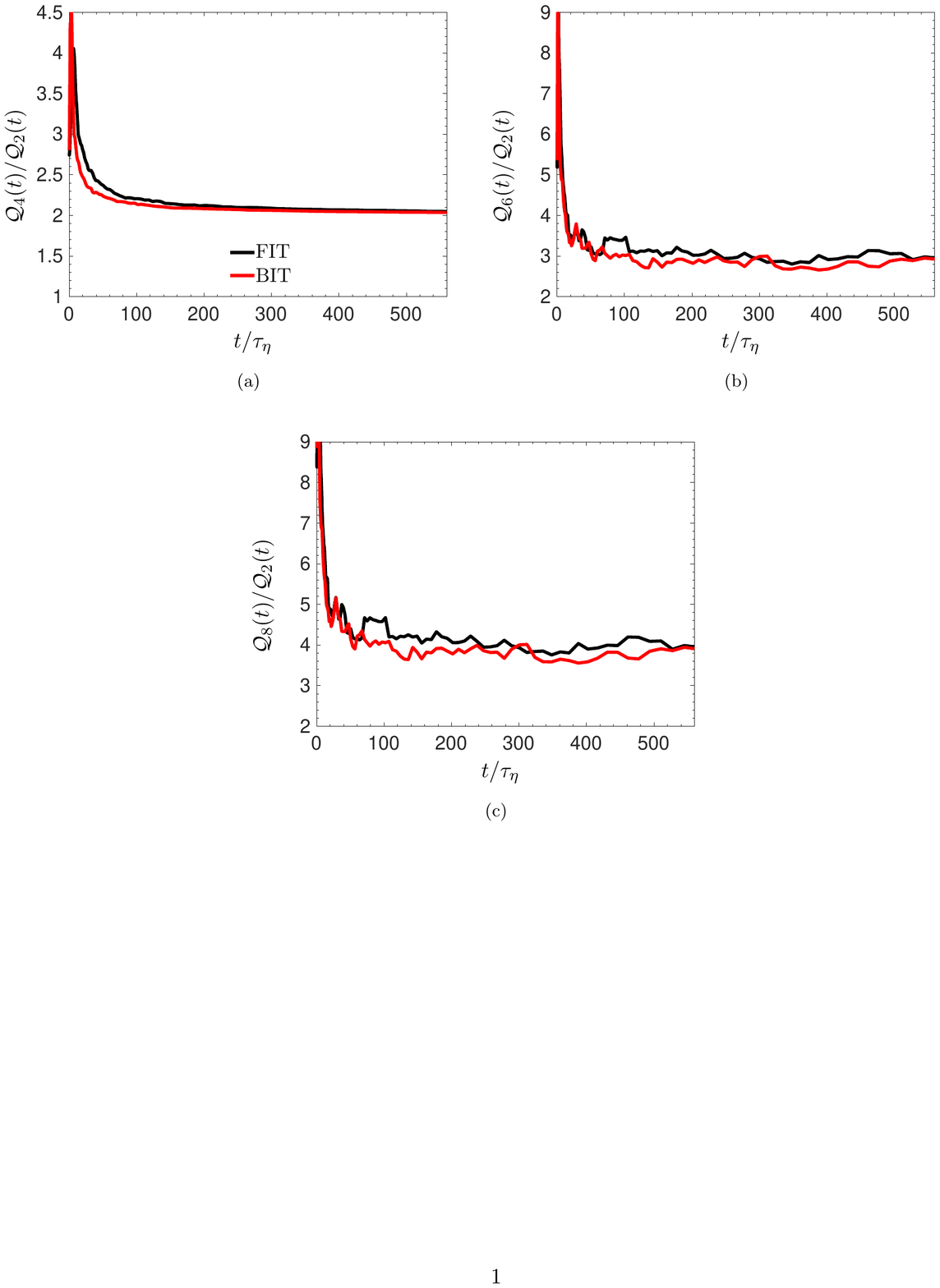}
	\end{overpic}}
	\caption{DNS data for (a) $\mathcal{Q}_4(t)/\mathcal{Q}_2(t)$, (b) $\mathcal{Q}_6(t)/\mathcal{Q}_2(t)$, and (c) $\mathcal{Q}_8(t)/\mathcal{Q}_6(t)$, where $\mathcal{Q}_N(t)=(\tau_\eta/t)\ln[\xi^{-N}\langle{r}^N(t)\rangle_\xi]$. The BIT results correspond to $\mathcal{Q}_N(-t)=(\tau_\eta/t)\ln[\xi^{-N}\langle{r}^N(-t)\rangle_\xi]$. If $\langle{r}^N( t)\rangle_\xi= \xi^N e^{N\mu^F t}$, then $\mathcal{Q}_{2+N}(t)/\mathcal{Q}_2(t)=(2+N)/2$.}
	\label{QN}
\end{figure}
\FloatBarrier
The results show that for $t\gtrsim 200\tau_\eta$, the FIT and BIT become approximately constant and remain so up to $t=560\tau_\eta$, which is the maximum simulation time. To the best of our knowledge, this is the first clear evidence for the exponential predictions $\langle{r}^N( t)\rangle_\xi= \xi^N e^{N\mu^F t}$ and $\langle{r}^N(-t)\rangle_\xi= \xi^N e^{-N\mu^B t}$ for fluid particle-pairs dispersing in the dissipation range of turbulence. However, these results also emphasize that the predictions $\langle{r}^N( t)\rangle_\xi= \xi^N e^{N\mu^F t}$ and $\langle{r}^N(-t)\rangle_\xi= \xi^N e^{-N\mu^B t}$ only apply at very long times in the dispersion process, and therefore great caution must be used when modeling fluid particle dispersion in the dissipation range using the exponential prediction. The results also illustrate why it has been so difficult previously to observe the exponential growth, since their initial separation has to be extremely small if their separation is to remain in the dissipation range for $t\gtrsim 200\tau_\eta$. Indeed, or results imply that in order to satisfy $r(t)\leq \eta$ at $t=200\tau_\eta$, we require $\xi/\eta\leq\mathcal{O}(10^{-22})$ for the FIT case, and $\xi/\eta\leq\mathcal{O}(10^{-33})$ for the BIT case. This makes it almost impossible to observe the long-time exponential growth of $\langle{r}^N(-t)\rangle_\xi$ in an experiment. 

As a further test, in Fig.~\ref{QN} we plot the ratio of the moments $\mathcal{Q}_{2+N}(t)/\mathcal{Q}_2(t)$. In the regime where $\langle{r}^N( t)\rangle_\xi= \xi^N e^{N\mu^F t}$ and $\langle{r}^N(-t)\rangle_\xi= \xi^N e^{-N\mu^B t}$, $\mathcal{Q}_{2+N}(t)/\mathcal{Q}_2(t)=\mathcal{Q}_{2+N}(-t)/\mathcal{Q}_2(-t)=(2+N)/2$. The results confirm this prediction very well, both FIT and BIT. It is also interesting to note that $\mathcal{Q}_{2+N}/\mathcal{Q}_2=(2+N)/2$ is satisfied even at times for which, as shown in Fig.~\ref{r2_r4}, $\langle{r}^N( t)\rangle_\xi= \xi^N e^{N\mu^F t}$ and $\langle{r}^N(-t)\rangle_\xi= \xi^N e^{-N\mu^B t}$ are not satisfied.

\begin{figure}[ht]
	\centering
	\vspace{-20mm}
	\hspace{0mm}{
		{\begin{overpic}
				[trim = 0mm 40mm 0mm 0mm,scale=0.38,clip,tics=20]{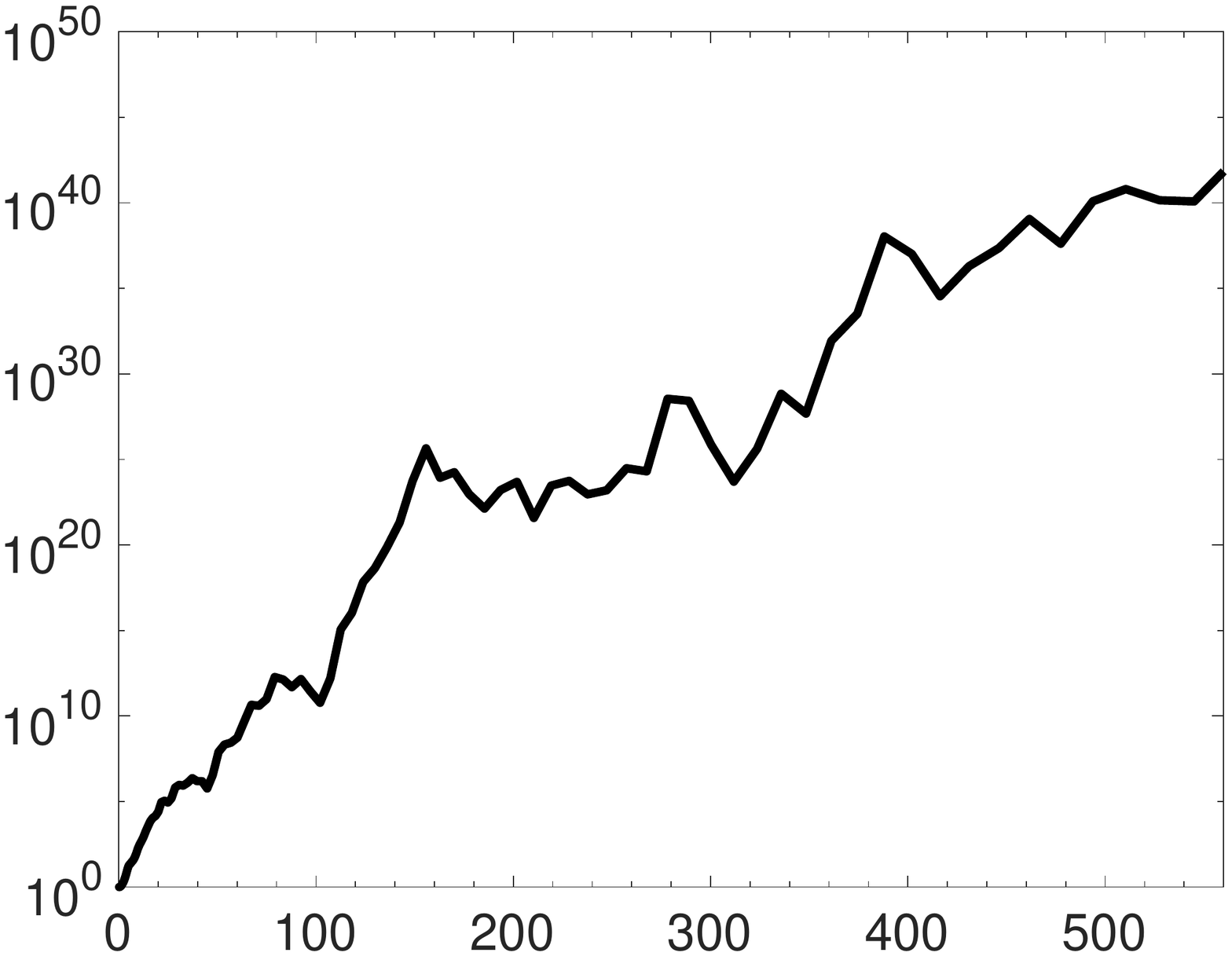}
				\put(115,8){$t/\tau_\eta$}
				\put(-20,70){\rotatebox{90}{$\langle{r}^2(-t)\rangle_\xi/\langle{r}^2(t)\rangle_\xi  $}}
		\end{overpic}}	
	}
	\caption{DNS data for $\langle{r}^2(-t)\rangle_\xi/\langle{r}^2(t)\rangle_\xi  $.}
	\label{Ratio}
\end{figure}
\FloatBarrier
In Fig.~\ref{Ratio} we plot the ratio of the BIT to FIT mean-square separation $\langle{r}^2(-t)\rangle_\xi/\langle{r}^2(t)\rangle_\xi$. In agreement with the discussion in \S\ref{LTR}, $\langle{r}^2(-t)\rangle_\xi/\langle{r}^2(t)\rangle_\xi$ grows (approximately) exponentially in time at long-times, such that the irreversibility of the dispersion is enormous. The higher-order moments $\langle{r}^N(-t)\rangle_\xi/\langle{r}^N(t)\rangle_\xi$ also exhibit the same behavior. As discussed earlier, this is in stark contrast to the behavior in the inertial range, where, if Richardson scaling holds,  $\langle{r}^2(-t)\rangle_\xi/\langle{r}^2(t)\rangle_\xi=\mathcal{O}(1)$ constant \cite{buaria15,bragg16}. 
\begin{figure}[ht]
	\centering
	\vspace{-30mm}
	{\begin{overpic}
		[trim = 10mm 165mm 0mm 0mm,scale=1,clip,tics=20]{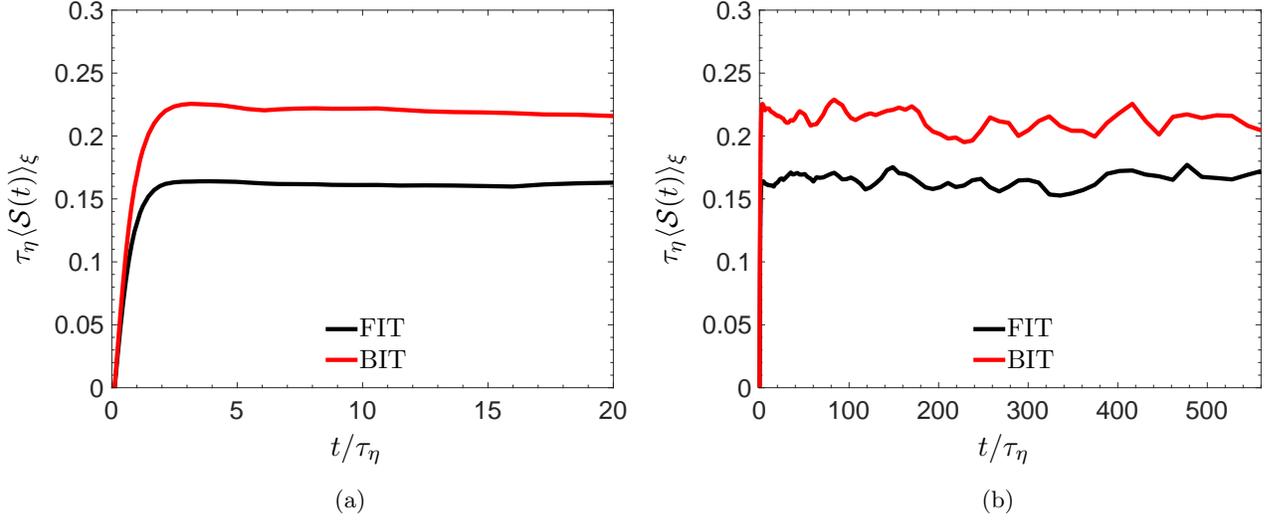}
	\end{overpic}}	
	\caption{DNS data for $\tau_\eta\langle\mathcal{S}( t)\rangle_\xi$, with plot (a) emphasizing the behavior at short times. For the BIT data we plot $-\tau_\eta\langle\mathcal{S}(-t)\rangle_\xi$ since $\langle\mathcal{S}(-t)\rangle_\xi\leq 0$. }
	\label{Lya}
\end{figure}
\FloatBarrier
Another quantity that was considered in \cite{gir90b,goto07} is
\begin{align}
\Big\langle \frac{d}{dt}\ln[r(t)]\Big\rangle_\xi\equiv \Big\langle \mathcal{S}(t)\Big\rangle_\xi.\label{Lya}
\end{align}
If $\langle\mathcal{S}(t)\rangle_\xi$ is constant it implies that on average the particles are separating exponentially. Our results for $\langle\mathcal{S}(t)\rangle_\xi$, both FIT and BIT, are shown in Fig.~\ref{Lya}, where for the BIT data we plot $-\tau_\eta\langle\mathcal{S}(-t)\rangle_\xi$ since $\langle\mathcal{S}(-t)\rangle_\xi\leq 0$. The results show, in agreement with those in \cite{gir90b,goto07}, that $\langle\mathcal{S}(t)\rangle_\xi$ does indeed become constant, even at relatively short-times, i.e. $t/\tau_\eta\geq \mathcal{O}(1)$. This seems in tension with our earlier results that show that exponential growth of $\langle r^N(t)\rangle_\xi$ is only observed for very large times, i.e. $t/\tau_\eta\gtrsim 200$. However, the difference is that \eqref{Lya} only depends, by definition, upon the mean of $\mathcal{S}$, whereas $\langle r^N(t)\rangle_\xi$ is affected by fluctuations in $\int_0^t\mathcal{S}(t')\,dt'$ about the mean value. Evidently, the impact of these fluctuations on $\langle r^N(t)\rangle_\xi$ persists for very long times, such that exponential growth of $\langle r^N(t)\rangle_\xi$ is only observed for $t/\tau_\eta\gtrsim 200$.
 
\begin{figure}[ht]
	\centering
	\vspace{-10mm}
	        {\begin{overpic}
		[trim = 0mm 50mm 0mm 30mm,scale=0.4,clip,tics=20]{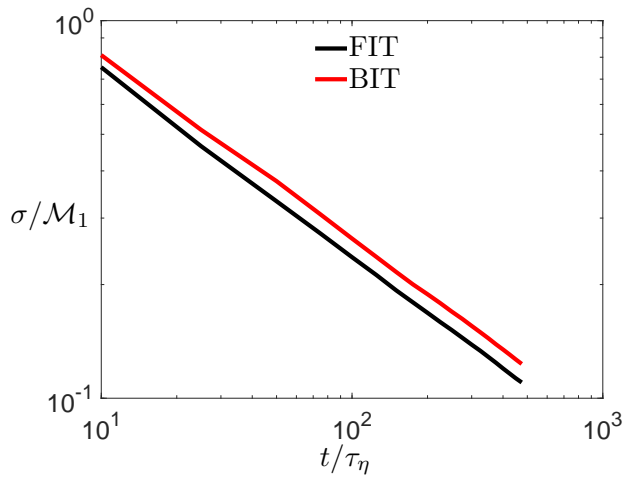}
		\put(115,8){$t/\tau_\eta$}
		\put(-2,100){{$\sigma/\mathcal{M}_1$}}
							\put(126,163){\small{FIT}}
			\put(126,151){\small{BIT}}
	\end{overpic}}			
	\caption{DNS data for $\sigma/\mathcal{M}_1$, where $\sigma\equiv \sqrt{(\mathcal{M}_2-\mathcal{M}_1^2)}$. The BIT results are $-\sigma/\mathcal{M}_1$ since $\mathcal{M}_1<1$ BIT.}
	\label{rmsM1}
\end{figure}

\begin{figure}[ht]
	\centering
	\vspace{-30mm}
	{\begin{overpic}
		[trim = 10mm 150mm 0mm 30mm,scale=1,clip,tics=20]{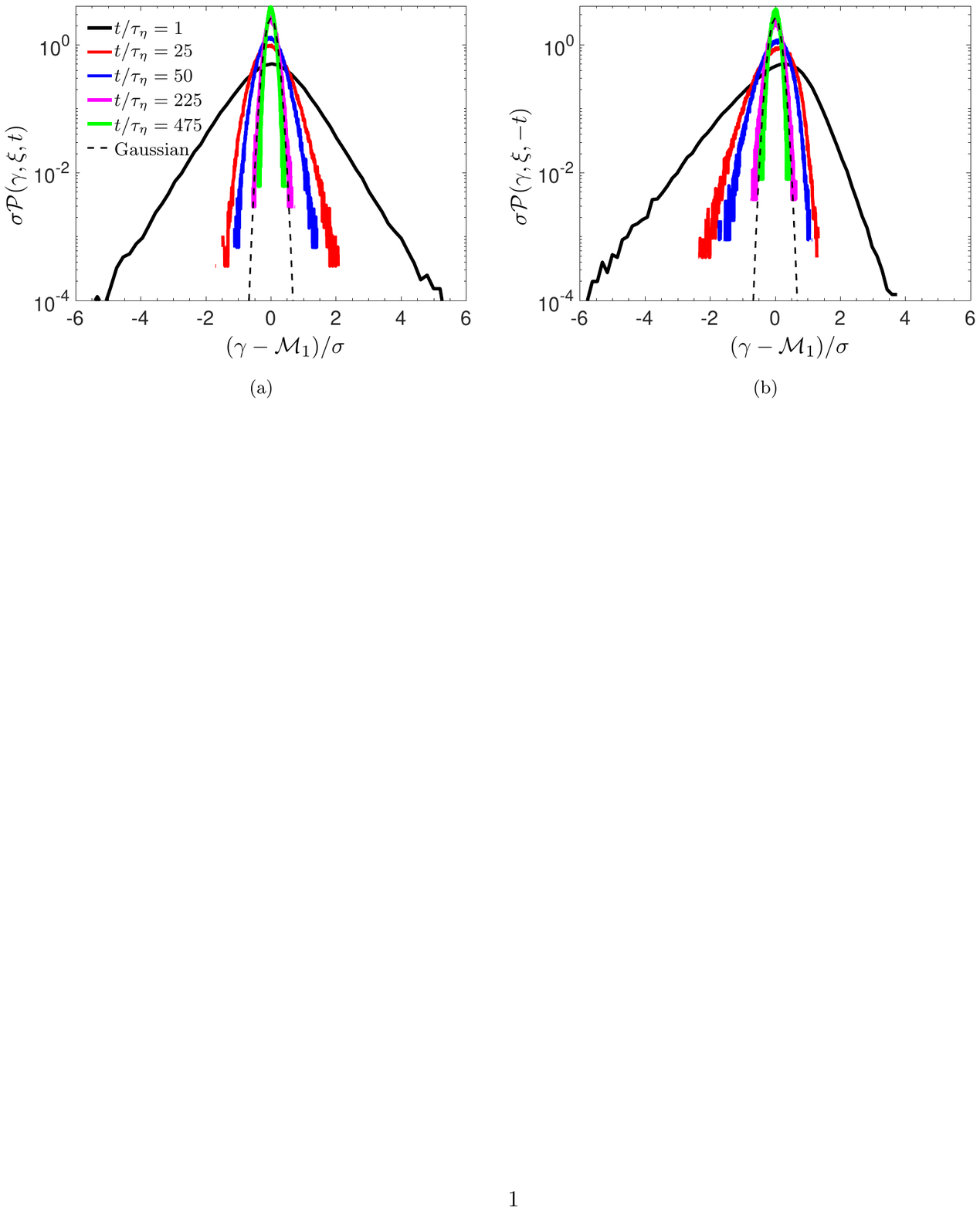}
	\end{overpic}}			
	\caption{DNS data for the PDFs (a) $ \mathcal{P}(\gamma,\xi,t)$, and (b) $ \mathcal{P}(\gamma,\xi,-t)$, plotted in standard form.}
	\label{PDF_strain}
\end{figure}

 To obtain more insight into the fluctuations of $\int_0^t\mathcal{S}(t')\,dt'$ and their impact upon $\langle r^N(t)\rangle_\xi$, we consider the Probability Density Function (PDF) of $\int_0^t\mathcal{S}(t')\,dt'$, namely
 \begin{align}
 \mathcal{P}(\gamma,\xi,t)\equiv\Bigg\langle \delta\Bigg(\int_0^t\mathcal{S}(t')\,dt' -\gamma  \Bigg)\Bigg\rangle_\xi,
 \end{align}
 whose moments are given by
 \begin{align}
 \mathcal{M}_N(\xi,t)\equiv \int_{\mathbb{R}} \mathcal{P}(\gamma,\xi,t)\gamma^N\,d\gamma.
 \end{align}
 The results in Fig~\ref{rmsM1} show the ratio of the r.m.s. $\sigma\equiv \sqrt{(\mathcal{M}_2-\mathcal{M}_1^2)}$ to the mean $\mathcal{M}_1$ both FIT and BIT. The results show that $\sigma/\mathcal{M}_1\propto t^{-1/2}$, as predicted by Batchelor \cite{batchelor52b}, implying that the growth of $\int_0^t\mathcal{S}(t')\,dt'$ is dominated by the mean value at long-times. This is why the low-order measure of dispersion $\langle (d/dt)\ln[r(t)]\rangle_\xi$ indicates exponential growth of $r(t)$, whereas the higher-order measures $\langle r^N(t)\rangle_\xi$, which are more sensitive to fluctuations in $\int_0^t\mathcal{S}(t')\,dt'$, do not exhibit exponential growth until much longer times, when the fluctuations about $\mathcal{M}_1$ become sufficiently small. The results also show that the fluctuations about the mean are stronger BIT than FIT. This is because the BIT separation of particle-pairs is dominated by $\mathcal{S}<0$, whereas the FIT separation of particle-pairs is dominated by $\mathcal{S}>0$, and because $\mathcal{S}$ has a negatively skewed PDF in 3D turbulence.
\FloatBarrier
 Finally, we recall from \S\ref{LTR} that in the regime $t\gg\tau_\eta$, it has been argued that $\langle r^N(t)\rangle_\xi$ can be predicted by appealing to the Central Limit Theorem (CLT), which implies that $ \mathcal{P}(\gamma,\xi,t)$ becomes Gaussian for $t\gg\tau_\eta$. Our results in Fig~\ref{PDF_strain} do indeed show that for $t/\tau_\eta\gtrsim 225 $,  $\mathcal{P}(\gamma,\xi,t)$ and $\mathcal{P}(\gamma,\xi,-t)$ become approximately Gaussian. However, closer inspection shows that $\mathcal{P}(\gamma,\xi,t)$ and $\mathcal{P}(\gamma,\xi,-t)$ actually become slightly sub-Gaussian, with tails that decay faster than a Gaussian. Possible reasons for this are that the CLT cannot apply exactly in the current context both because $\mathcal{S}$ is differentiable, and therefore $\int_0^t\mathcal{S}(t')\,dt'$ can never strictly be considered the sum of a large number of independent numbers, and second, because the CLT does not account for extreme events in $\mathcal{S}$.
\FloatBarrier

\section{Conclusions}
In this paper we have considered how the statistical moments of the separation between two fluid particles, $\langle r^N(t)\rangle_\xi$, grow with time when $r(t)$ lies in the dissipation range of turbulence. For short-times, $\langle r^N(t)\rangle_\xi$ grows as a power law in time $t$, however, at long times, there are theoretical arguments for exponential growth of $\langle r^N(t)\rangle_\xi$. Although there have been claims in the literature to observe the exponential growth of $\langle r^N(t)\rangle_\xi$ in the dissipation range, these claims are problematic, especially since several of them claim to observe the exponential growth in the short-time regime, where on theoretical grounds it is not supposed to occur. Therefore, in order to attempt to settle the question, we have conducted Direct Numerical Simulations (DNS) to compute $\langle r^N(t)\rangle_\xi$ over very long times, $t\leq 560\tau_\eta$. The results show that if the initial separation between the particles is infinitesimal, the moments of the particle separation first grow as power laws in time, but we then observe, for the first time, convincing evidence that at sufficiently long times the moments grow exponentially. However, this exponential growth is only observed after extremely long times $\gtrsim 200\tau_\eta$, where $\tau_\eta$ is the Kolmogorov timescale. We computed the statistics of the strain-rate along the particle trajectories
 and showed that the deviations from exponential growth are due to strong fluctuations in the strain-rate about its mean value, which affects the moments of the particle separation for very long times into the dispersion process. We also consider the Backward-in-Time (BIT) separation of the particles, and observe that it too grows exponentially in the long-time regime. However, a dramatic consequence of the exponential separation is that at long-times the difference between the rate of the particle separation Forward-in-Time (FIT) and BIT grows exponentially in time, leading to incredibly strong irreversibility in the dispersion. This is in striking contrast to the irreversibility of their relative dispersion in the inertial range, where the difference between FIT and BIT is approximately constant in time.

\bibliography{refs_co12}

\end{document}